\documentclass[prb,reprint,twocolumn,showpacs]{revtex4}
\usepackage{amsmath,amssymb,graphicx,xcolor,units,hyperref}

\newcommand{\dd}[1]{\mathrm{d}#1\,}
\newcommand{\avg}[1]{\langle{#1}\rangle}
\renewcommand{\Re}{\mathop{\mathrm{Re}}}
\renewcommand{\Im}{\mathop{\mathrm{Im}}}
\newcommand{\Tr}{\mathop{\mathrm{Tr}}}

\DeclareMathOperator{\arccosh}{arccosh}
\DeclareMathOperator{\arctanh}{arctanh}
\DeclareMathOperator{\arcsinh}{arcsinh}

\newcommand\varpm{\mathbin{\vcenter{\hbox{%
  \oalign{\hfil$\scriptstyle+$\hfil\cr
          \noalign{\kern-.3ex}
          $\scriptscriptstyle({-})$\cr}%
}}}}
\newcommand\varmp{\mathbin{\vcenter{\hbox{%
  \oalign{$\scriptstyle({+})$\cr
          \noalign{\kern-.3ex}
          \hfil$\scriptscriptstyle-$\hfil\cr}%
}}}}

\begin{document}

\title{Fluctuation of heat current in Josephson junctions}

\author{P. Virtanen}
\affiliation{O.V. Lounasmaa Laboratory, Aalto University,
  P.O. Box 15100,FI-00076 AALTO, Finland}

\author{F. Giazotto}
\affiliation{NEST, Istituto Nanoscienze-CNR and Scuola Normale Superiore, I-56127 Pisa, Italy}

\date{\today}


\begin{abstract}
  We discuss the statistics of heat current between two
  superconductors at different temperatures connected by a generic
  weak link.  As the electronic heat in superconductors is carried by
  Bogoliubov quasiparticles, the heat transport fluctuations follow
  the Levitov--Lesovik relation. We identify the energy-dependent
  quasiparticle transmission probabilities and discuss the resulting
  probability density and fluctuation relations of the heat current.
  We consider multichannel junctions, and find that heat transport
  in diffusive junctions is unique in that its statistics is
  independent of the phase difference between the superconductors.
  Curiously, phase dependence reappears if phase coherence is partially
  broken.
\end{abstract}

\maketitle

\section{Introduction}

Heat transport through junctions between superconductors is
significantly affected by superconductivity.  \cite{maki1965} In
tunnel Josephson junctions, superconducting phase coherence manifests
as a component of the thermal conductance that oscillates with the
phase difference between the superconductors,
\cite{guttman1997-pdt,guttman1998-ieh,zhao2003-phase,zhao2004-htt} a
prediction which was confirmed by recent
experiments. \cite{giazotto2012-jhi,martinez2014-qdt} In general, both the sign and the
magnitude of the oscillations depend on the transparency of the
junction in question. \cite{zhao2003-phase}

Previous studies have largely concentrated on the ensemble average
value of the heat currents.  However, in reality the heat current
driven by a temperature difference through a superconducting junction
is not constant in time, but fluctuates. When mesoscopic systems are
considered, this can lead to fluctuations in other quantities --- such
as the energy stored on a small metal island --- and eventually, in
measurable observables, such as charge current
\cite{heikkila2009-stf,laakso2012-ttf,laakso2012-mng} or temperature
measured by a generic temperature probe \cite{utsumi2014-fth}. In
addition to the theoretical question on how the coherent physics of
phase differences in superconducting order parameters manifest in
statistical properties of heat transport, questions on fluctuations
can also be of interest in systems that utilize mesoscopic
superconductors in a nonequilibrium settings for example for radiation
detection \cite{giazotto2006-omi}.

\begin{figure}
  \includegraphics{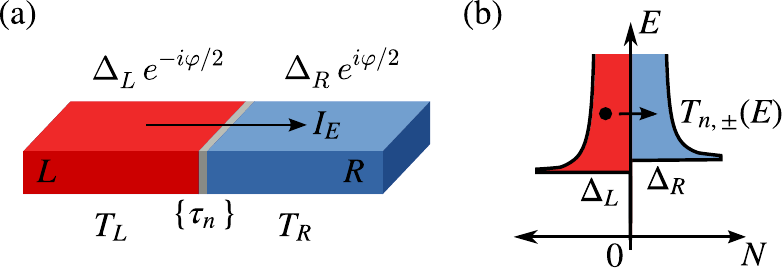}
  \caption{\label{fig:setup} 
    (a) Heat current $I_E$ flows from
    superconductor $L$ to another superconductor $R$ coupled to it,
    driven by a difference in the temperatures $T_L>T_R$.  
    The heat current is modulated by the phase difference $\varphi$
    between the order parameters $\Delta_L$, $\Delta_R$ of the superconductors.
    The Josephson junction connecting the two is described by the set of
    (spin-independent) transmission eigenvalues $\{\tau_n\}$ of
    the normal-state scattering matrix of the interface.  (b) 
    The excitations that carry heat in
    superconductors are Bogoliubov quasiparticles, whose density of
    states $N(E)$ is illustrated. Heat current and its fluctuation
    statistics is fully determined by their energy-dependent
    transmission probabilities $T_{n\pm}(E)$. Each normal-state
    quantum channel ($\tau_n$) splits into inequivalent
    particle-hole transmission channels ($\pm$) due to
    superconductivity.
  }
\end{figure}

In this work, we find the full statistics of temperature-driven heat
transport in a Josephson junction of arbitrary transparency, as
illustrated in Fig.~\ref{fig:setup}. Previously, the energy transport
statistics has been discussed in the tunneling limit,
\cite{golubev2013-htt} and we recover these results as a special case.
We separate the energy transport to elementary quasiparticle transport
events.  In agreement with the Levitov--Lesovik formula,
\cite{levitov1993-cdi,levitov96} the elementary quasiparticle
transmission probabilities are governed by the total transmission
eigenvalues $T_{n\pm}(E)$ [Eq.~\eqref{eq:tprob}] of the Bogoliubov--de
Gennes scattering problem of the interface.  We compute the heat
current noise and discuss its parameter dependence in single-channel
junctions. We derive results corresponding to dirty-interface
\cite{schep97} and diffusive \cite{dorokhov1984} multichannel
junctions. The heat statistics in the diffusive limit turns out to
have no phase dependence, except in the presence of inelastic effects.

\section{Model}

We consider heat transport between two superconducting terminals that
are connected by a generic contact described by the transmission
eigenvalues $\{\tau_j\}$.  The terminals are assumed to lie at different
temperatures. We make use of the Keldysh-Nambu Green function
formulation for transport in superconducting structures,
\cite{larkin1986-ns,belzig1999-qgf,usadel1970-gde} and the
quasiclassical boundary condition description of a weak link between
bulk superconductors in the diffusive limit.
\cite{zaitsev84,nazarov99}

At equilibrium, electrons inside a superconducting terminal at
temperature $T$ with superconducting gap $\Delta$ are described by the
quasiclassical equilibrium Green function $\check{g}_0(E)$,
\begin{align}
  \label{eq:g-equilibrium}
  \check{g}_0 
  &= 
  \begin{pmatrix}
    \hat{g}^R_0 & (\hat{g}^R_0-\hat{g}^A_0)h_0 \\ 0 & \hat{g}^A_0
  \end{pmatrix}
  \,,
  \\
  \hat{g}^R 
  &= 
  \begin{pmatrix}
    \frac{E}{\sqrt{E^2 - |\Delta|^2}} & \frac{\Delta}{\sqrt{E^2 - |\Delta|^2}} \\     
    -\frac{\Delta^*}{\sqrt{E^2 - |\Delta|^2}} & -\frac{E}{\sqrt{E^2 - |\Delta|^2}}
  \end{pmatrix}
  \,,
\end{align}
and $\hat{g}^A=-\hat{\tau}_3(\hat{g}^R)^\dagger\hat{\tau}_3$, where
$\hat{\tau}_3$ is the third spin matrix in the Nambu space.
Temperature enters in the equilibrium distribution function
$h_0=\tanh\frac{E}{2T}$.  Presence of sub-gap states in
superconductors can be taken into account via a Dynes parameter, by
replacing $E\mapsto{}E\pm{}i\Gamma/2$ in $g^{R/A}$, where $\Gamma$ is
a relaxation rate due to e.g. electron-phonon
or other interactions.

Statistics of heat flow can be conveniently described via the
two-point generating function
\cite{kindermann2004-sht,esposito2009-nff}
\begin{align}
  \label{eq:w-definition}
  W_\alpha(u,t) = \Tr[e^{i u H_\alpha} U(t) e^{-iuH_\alpha} \rho(0) U(t)^\dagger]
  \,,
\end{align}
where $\alpha=L,R$ indicates the terminal whose internal energy is
counted, and $H_\alpha$ are the BCS Hamiltonians of the
superconducting terminals.  These functions can be computed using the
Keldysh approach of
Refs.~\onlinecite{nazarov99b,belzig01,kindermann2004-sht}, as
follows. 
Differentiating Eq.~\eqref{eq:w-definition} we obtain,
\begin{align}
  \label{eq:energy-flow}
  \partial_u W(u,t)
  =
  i\avg{H_\alpha(t) - H_\alpha(0)}_u W(u,t)
  \,,
\end{align}
where $\avg{X}_u=\Tr[X U_+ \rho(0)
U_-^\dagger]/\Tr[U_+\rho(0)U_-^\dagger]$ is an expectation value
computed with modified time evolution operators
$U_\pm=e^{\pm{}iuH_\alpha/2}Ue^{\mp{}iuH_\alpha/2}$ including the
counting field with differing signs on different Keldysh
branches. This results only to time shifts in the interaction picture Green
function of lead $\alpha$, as the energy counting factor has the same
form as time evolution. \cite{kindermann2004-sht}
Transforming to the Green function representation \cite{rammer86} used
above, the time shifts are represented by
\begin{align}
  \check{g}_\alpha(E,u)
  &=
  e^{iuE\check{\sigma}_1/2}\check{g}_{\alpha}(E)e^{-iuE\check{\sigma}_1/2}
  \,.
\end{align}
Computing the expectation value in Eq.~\eqref{eq:energy-flow} via the
quasiclassical boundary condition approach,
\cite{zaitsev84,nazarov99} and integrating in $u$ results to the
well-known action of superconducting contacts
\cite{nazarov99b,belzig01,kindermann2004-sht,snyman2008-kam}
\begin{align}
  \label{eq:generating-function}
  \ln W_R(u,t)
  =
  \frac{1}{2}
  \sum_n
  \Tr
  \ln
  \Bigl[1 + \frac{\tau_n}{4}([\check{g}_L,\check{g}_R(u)]_+ - 2)\Bigr]
  +
  C
  \,,
\end{align}
where $C$ is a normalization constant, and $\Tr$ includes energy
integration in addition to Keldysh-Nambu matrix trace.

\section{Generating function}
\label{sec:generating-function}

An important difference in energy transport compared to charge
statistics follows from the fact that an Andreev reflection does not
transfer energy. In the present formulation, this is visible in the
fact that ($\Gamma\to0^+$)
\begin{gather}
  \label{eq:gr-stationary}
  \check{g}(E,u)
  =
  \hat{g}^R(E)
  \otimes
  \begin{cases}
    1 \,, 
    & 
    |E|<\Delta
    \,,
    \\
    e^{\frac{iuE\check{\sigma}_1}{2}}
    \begin{pmatrix}
      1 & 2h \\ 
      0 & -1
    \end{pmatrix}
    e^{\frac{-iuE\check{\sigma}_1}{2}}
    \,,
    & 
    |E|>\Delta\,.
  \end{cases}
\end{gather}
There is no energy transfer at sub-gap energies, where there are no
quasiparticles, assuming no broadening in the spectrum
of the superconductors.

The generating function can be found by direct substitutions into
Eq.~\eqref{eq:generating-function}. It is however useful to make use
of $\check{g}_{L/R}^2=1$ and rewrite
\begin{align}
  1 + \frac{\tau_n}{4}([\check{g}_L,\check{g}_R(u)]_+ - 2)
  =
  \frac{
    [q_n + \check{g}_L\check{g}_R(u)][q_n + \check{g}_R(u)\check{g}_L]
  }{
    (1 + q_n)^2
  }
  \,,
\end{align}
where $q_n=-1+2/\tau_n+2\sqrt{1-\tau_n}/\tau_n$ are the eigenvalues of
the hermitian square of the corresponding transfer matrix,
\cite{nazarov99b,beenakker97} so that
\begin{align}
  \label{eq:generating-function-lndet}
  \ln W_R(u)
  =
  \sum_n
  \ln\det[q_n + \check{g}_L\check{g}_R(u)]
  +
  C'
  \,,
\end{align}
where $C'$ is a normalization constant.

The product structure $\check{g}=\hat{g}^R\otimes\check{V}$ of
Eq.~\eqref{eq:gr-stationary} implies that the Keldysh and Nambu
components can be diagonalized separately. This yields
\begin{align}
  \label{eq:generating-function-expl}
  \ln W_R(u)
  =
  2t_0
  \int_{\max(\Delta_L,\Delta_R)}^\infty\frac{\dd{E}}{2\pi}
  \sum_n
  \sum_{\alpha\beta=\pm1}
  \ln
  \frac{\mu^\alpha + q_n\lambda^{\beta}}{1 + q_n \lambda^\beta}
  \,,
\end{align}
where $t_0$ is the measurement time, and $\{\lambda,1/\lambda\}$ and
$\{\mu,1/\mu\}$ are the eigenvalues of $\hat{g}^R_L\hat{g}^R_R$ and
$\check{V}_L\check{V}_R(u)$, respectively.

Solving the eigenvalue problems, Eq.~\eqref{eq:generating-function-expl}
can be rewritten in the form of a characteristic function of a
multinomial distribution:
\begin{align}
  \label{eq:stationary-statistics}
  \ln W_R(u)
  &=
  2t_0
  \int_{0}^\infty\frac{\dd{E}}{2\pi}
  \sum_n
  \sum_{\beta=\pm1}
  \ln
  \sum_{k}
  e^{ikuE}p_{k,n,\beta}(E)
  \,,
\end{align}
with the event probabilities
\begin{align}
  p_{k,n,\beta}(E)
  =
  \begin{cases}
    T_{n\beta}(E)[1 - f_R(E)]f_L(E)\,, & k=+1\,,
    \\
    T_{n\beta}(E)[1 - f_L(E)]f_R(E)\,, & k=-1\,,
    \\
    1 - p_{+1,n,\beta}(E) - p_{-1,n,\beta}(E)\,, & k=0\,,
    \\
    0\,, & \text{otherwise}\,.
  \end{cases}
\end{align}
Here, $f_{L/R}$ are the electron Fermi distribution functions in the
left and right terminals, and
\begin{subequations}
\label{eq:tprob}
\begin{gather}
  T_{n\beta}(E)
  =
  \theta(E^2-|\Delta_L|^2)
  \theta(E^2-|\Delta_R|^2)
  \frac{
    4 \lambda^{\beta} q_n
  }{
    (1 + \lambda^{\beta} q_n)^2
  }
  \,,
  \\
  \lambda
  =
  \exp
  \arccosh
  \frac{
    E^2 - |\Delta_L||\Delta_R|\cos\varphi
  }{
    \sqrt{E^2 - |\Delta_L|^2}\sqrt{E^2 - |\Delta_R|^2}
  }
  \,.
\end{gather}
\end{subequations}
Here, $\varphi$ is the phase difference between the order parameters
of the two superconductors.  The generating function describes events
where a quasiparticle at energy $E>\Delta$ attempts to move from the
left to the right ($k=+1$) or from the right to the left ($k=-1$). The
probability that such a transmission event succeeds depends both on
the transparency of the transmission channel ($q_n$, $\tau_n$), and on
superconductivity ($\lambda$).

The probabilities $T_{n,\pm}$ are the transmission eigenvalues of a
Bogoliubov--de Gennes (BdG) scattering problem. The corresponding
scattering matrix (for $\Delta_L=\Delta_R$) can be found in
Ref.~\onlinecite{beenakker1991-ulc}. Direct evaluation of the
transmission eigenvalues using the results there
gives
\begin{align}
  \mathrm{eig}\,
  t t^\dagger
  =
  \{T_{n,\pm}\}
  \,,
\end{align}
which coincides with Eq.~\eqref{eq:tprob} above.  A similar connection
can be made to well-known results in N/S junctions,
\cite{blonder1982-tfm} ($\Delta_L=0$, $\Delta_R=\Delta>0$), after
identifying the barrier reflectivity in
Ref.~\onlinecite{blonder1982-tfm} as $Z = (q-1)/\sqrt{4q}$.  That the
energy statistics is related to these scattering matrices stems from
the fact that the counting statistics of Bogoliubov quasiparticles,
which carry all of the electronic energy current, must follow the
well-known Levitov--Lesovik result \cite{levitov1993-cdi,levitov96}.

\begin{figure}
  \includegraphics{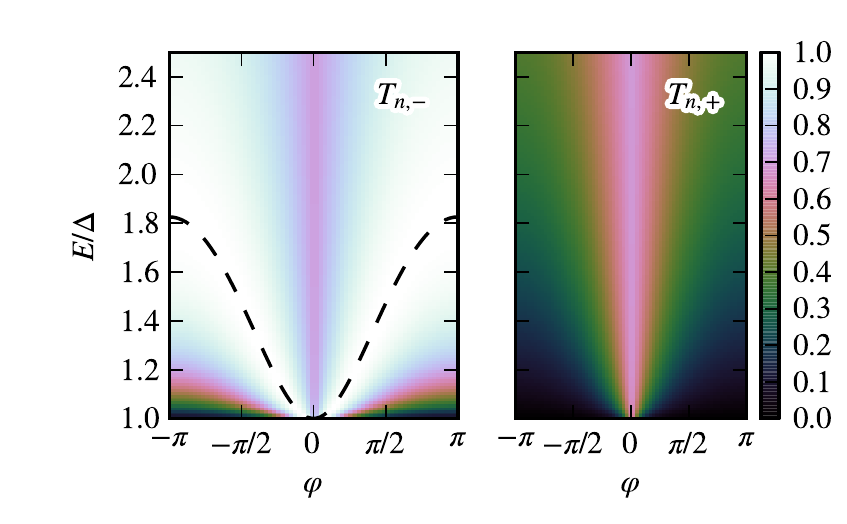}
  \caption{\label{fig:Tbetan}
    Transmission eigenvalues $T_{n,\pm}(E)$ 
    calculated for $\tau_n=0.7$ and $\Delta_L=\Delta_R=\Delta$.
    Location of the resonance~\eqref{eq:transmission-resonance}
    is indicated with a dashed line.
    At $\varphi=0$, $T_{n,\pm}=\tau_n$.
  }
\end{figure}

Superconductivity has a significant impact on the transmission
eigenvalues. In the normal state, they are simply
$T_{n\beta}=4q_n/(1+q_n)^2=\tau_n$ independent of the particle-hole
channel index $\beta=\pm1$. In the superconducting state, they deviate
significantly from this, as illustrated in Fig.~\ref{fig:Tbetan}. In
particular, there is a transmission resonance in one of the channels:
\begin{align}
  \label{eq:transmission-resonance}
  T_{n,-}(E_{\rm res})=1\,,\quad
  E_{\rm res} = \pm\Delta\sqrt{1 + \frac{\tau_n}{1-\tau_n}\sin^2\frac{\varphi}{2}}
  \,.
\end{align}
The above result applies for $\Delta_L=\Delta_R$, but the resonance
appears also for $\Delta_L\ne\Delta_R$. This has a large effect
especially for junctions whose normal-state transparency is small, in
which a large part of the total heat current is carried by the
resonance. \cite{zhao2003-phase}


\subsection{Breaking phase coherence}

We can extend the above results to include broadening of the density
of states in the superconductors, by adding a Dynes parameter
$\Gamma>0$. Also in this case, the final generating function obtains
the form~\eqref{eq:stationary-statistics}. As the
factorization~\eqref{eq:gr-stationary} does not apply, the expressions
for the transmission eigenvalues $T_{n\beta}(E)$ need to be extracted
from the determinant in Eq.~\eqref{eq:generating-function-lndet}.

\begin{widetext}
Straightforward calculation yields for $T_{n,\pm}(E)$
the indirect expressions
\begin{subequations}
\label{eq:tfull}
\begin{align}
  T_{n,+}T_{n,-}
  &=
  \frac{
    16 q_n^2 \cos^2(\Im\theta_L)\cos^2(\Im\theta_R)
  }{
    \left|
      1
      +
      q_n^2
      +
      2q_n
      \cosh\theta_L
      \cosh\theta_R
      -
      2 q_n
      \sinh\theta_L
      \sinh\theta_R
      \cos\varphi
    \right|^2
  }
  \,,
  \\
  \frac{1}{T_{n,+}} + \frac{1}{T_{n,-}}
  &=
  1
  +
  \tan(\Im\theta_L) \tan(\Im\theta_R) \cos\varphi 
+
  \frac{
    (q_n^2+1)
    [
    \cosh(\Re\theta_L) 
    \cosh(\Re\theta_R)
    -
    \sinh(\Re\theta_L) 
    \sinh(\Re\theta_R)
    \cos\varphi
    ]
  }{
    2 q_n \cos(\Im\theta_L) \cos(\Im\theta_R) 
  }
  \,,
\end{align}
\end{subequations}
where $\theta_{L/R} = \arctanh\frac{\Delta_{L/R}}{E + i(\Gamma/2)}$.
The above formulas reduce to Eq.~\eqref{eq:tprob} for $\Gamma\to0^+$.
For $\Gamma>0$, the results cannot however be written in terms of a
single $\lambda$ as in Eq.~\eqref{eq:tprob}.
\end{widetext}

\begin{figure}
  \includegraphics{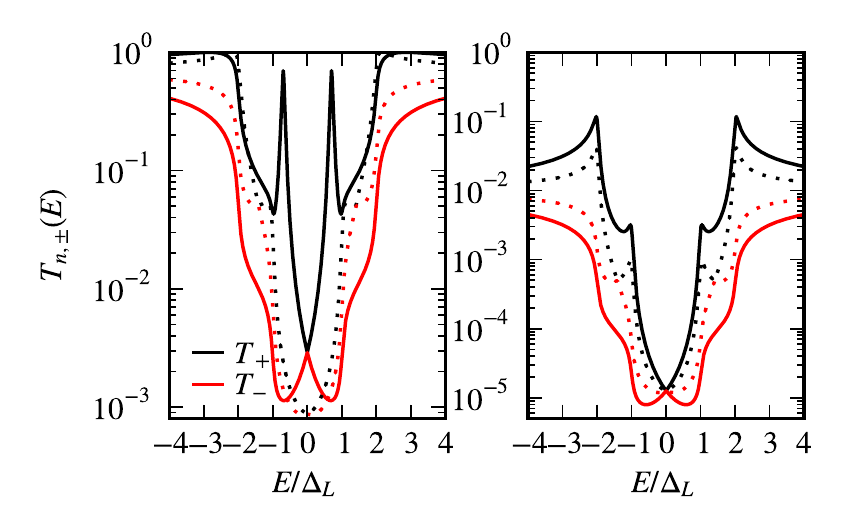}
  \caption{
    \label{fig:Tbetan-Gamma}
    Transmission eigenvalues $T_{n,\pm}(E)$ at $\varphi=\pi$ (solid) and $\varphi=0$ (dotted)
    for $\Gamma=0.1\Delta$, and $\tau_n=0.7$ (left panel) or $\tau_n=0.01$ (right panel).
    Here, $\Delta_R=2\Delta_L$.
  }
\end{figure}

The eigenvalues $T_{n,\pm}(E)$ are plotted for $\Gamma>0$ in
Fig.~\ref{fig:Tbetan-Gamma}. In general, a finite $\Gamma$ induces a
small number of states inside the gaps of the superconductors, so that
the junction is transparent for heat transport also at sub-gap
energies.  Similarly as for above-gap transport, the transmission
through the two channels $\pm$ can differ significantly. Moreover, we
can observe that sharp sub-gap resonances appear in one of the two
channels --- these are associated with the Andreev bound states and
are most prominent in transparent junctions.  Moreover, we note that
the phase dependence of sub-gap heat transport in high-transparency
channels (left panel) can be substantial, whereas it is not as
prominent at low transparencies (right panel).

\section{Heat flow statistics}

\begin{figure}
  \includegraphics{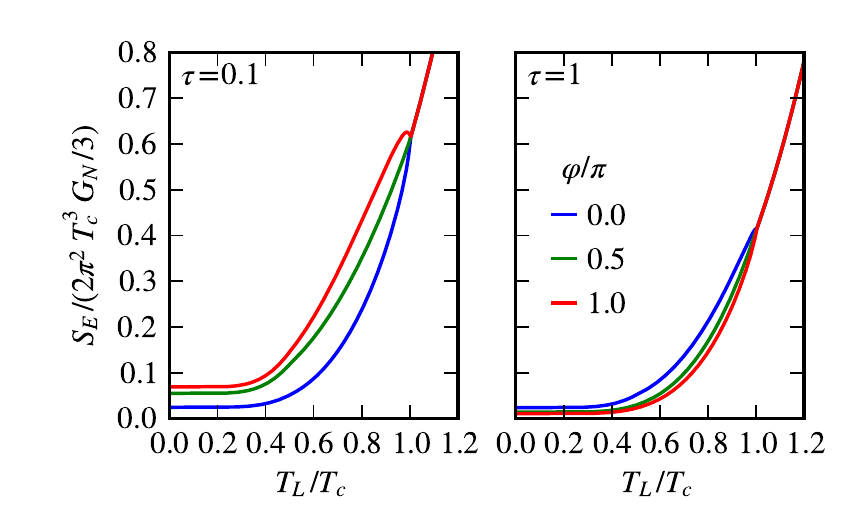}
  \caption{\label{fig:SE} 
    Heat current noise $S_E$ as a function of
    temperature $T_L$, for different values of $\varphi$ 
    and $\tau_n$ (single channel),
    normalized to its equilibrium normal-state value at $T=T_c$.
    The temperature $T_R=T_c/2$ is kept fixed.
    Here, $T_c$ is the BCS critical temperature, and we take the temperature
    dependence of the energy gaps $\Delta_L(T_L)$, $\Delta_R(T_R)$
    into account via the BCS relation, taking $\Delta_L(0)=\Delta_R(0)$.
  }
\end{figure}

Let us now compute the first moments of the heat statistics.  Direct
differentiation of Eq.~\eqref{eq:stationary-statistics} with respect
to $u$ yields the average heat current,
\begin{align}
  \label{eq:heat-current}
  I_E
  =
  \sum_n
  \int_{-\infty}^\infty
  \frac{\dd{E}}{2\pi}
  E
  [T_{n,+}(E) + T_{n,-}(E)]
  [f_L(E) - f_R(E)]
  \,,
  \\
  \sum_\pm
  T_{n\pm}
  =
  \frac{
    2\tau_n(E^2-\Delta^2)[E^2-\Delta^2\cos^2\frac{\varphi}{2} + r_n\Delta^2\sin^2\frac{\varphi}{2}]
  }{
    (E^2 - \Delta^2(1 - \tau_n\sin^2\frac{\varphi}{2}))^2
  }
  \,,
\end{align}
where $r_n=1-\tau_n$.  Naturally, this result coincides exactly with
that found in Ref.~\onlinecite{zhao2003-phase}. Note, however, that the
coefficients ${\cal D}_{ee}$, ${\cal D}_{he}$ defined in
Ref.~\onlinecite{zhao2003-phase} do not coincide with $T_{n\pm}$, even
though the sums do.

Taking the second derivative, we find the heat current zero-frequency
noise
\begin{align}
  S_E
  &=
  \sum_n\sum_{\beta=\pm1}
  \int_{-\infty}^\infty\frac{\dd{E}}{2\pi}
  E^2
  T_{n\beta}
  \Bigl\{
  f_L (1 - f_R) + f_R(1 - f_L)
  \\\notag
  &\qquad
  +
  T_{n\beta} (f_L - f_R)^2
  \Bigr\}
  \,.
\end{align}
In the tunneling limit, this result coincides with that found in
Ref.~\onlinecite{golubev2013-htt}. The behavior of the heat current
noise away from equilibrium is shown in Fig.~\ref{fig:SE}. We
emphasize the difference of $S_E$ between the opaque (left panel) and
transparent junction limit (right panel). In particular, in the former
case, the heat current noise is minimized for $\varphi=0$ whereas in
the latter junction $S_E$ is minimized for $\varphi=\pi$.  This
behavior resembles the one of the thermal conductance of a
temperature-biased Josephson weak-link, as predicted in
Refs.~\onlinecite{zhao2003-phase,zhao2004-htt}.

\begin{figure}
  \includegraphics{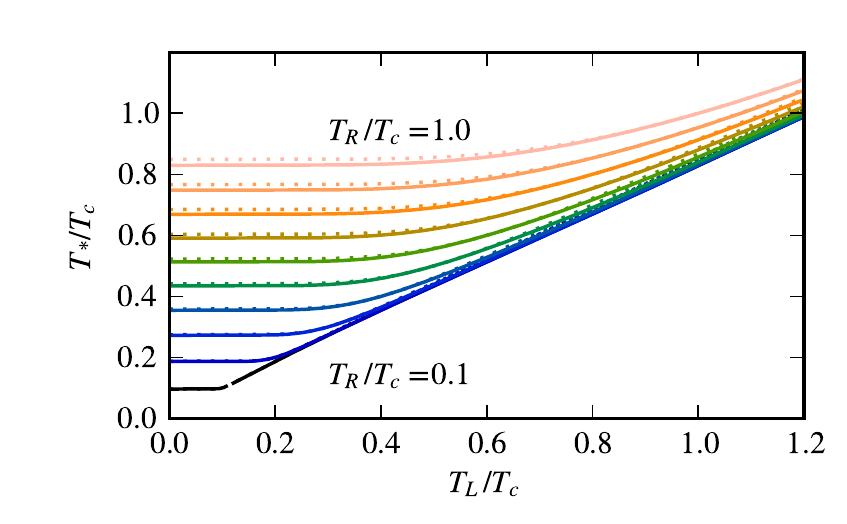}
  \caption{\label{fig:Tstar}
    Temperature $T_*(T_L,T_R)$ at which $S_E(T_L,T_R) = 2 T_*^2G_{\rm th}(T_*)$.
    $T_R$ sweeps from $0.1 T_c$ to $T_c$ in steps of $0.1T_c$.
    Shown for a single-channel system with $\tau=0.1$, $\varphi=\pi$ (solid)
    and $\tau=0.9$, $\varphi=\pi/4$ (dashed).
  }
\end{figure}

At equilibrium ($T_L=T_R=T$), the heat current noise obeys the
well-known fluctuation relation $S_E=2G_{th}T^2$ that connects it to
the thermal conductance $G_{th}=\dd{I_E}/\dd{T_R}\rvert_{T_L=T_R}$.
Conversely, we can define an effective temperature
$T_*(T_L,T_R,\{\tau_n\},\varphi)$ such that $S_E = 2T_*^2G_{\rm
  th}(T_*)$. Such a temperature plot is shown in
Fig.~\ref{fig:Tstar}. The result is fairly insensitive to the values
of $\tau_n$ and $\varphi$, and the results fall nearly on the same
curves for different values of these parameters.  At low temperatures,
the result converges towards $T_*=\max(T_L,T_R)$.


More generally, the generating function obeys the fluctuation relation
\cite{esposito2009-nff}
\begin{align}
  W_R(u)
  =
  W_R(-u + iT_L^{-1} - iT_R^{-1})
  \,.
\end{align}
For the probability distribution of transferring energy $\varepsilon$ out of
terminal $R$ in time $t_0$ this implies
\begin{align}
  \frac{P_R(\varepsilon,t_0)}{P_R(-\varepsilon,t_0)} = e^{\varepsilon/T_R}e^{-\varepsilon/T_L}
  \,,
  \quad
  t_0\to\infty
  \,.
\end{align}
This fluctuation relation is independent of the channel transmissions
$T_{n\beta}$, and therefore does not contain information about
superconductivity.

\subsection{Diffusive junctions}

It is possible to average the above generating functions over
known distributions of transmission eigenvalues $\{\tau_n\}$, to
obtain results for certain types of multichannel junctions.  Let us in
particular consider the transmission eigenvalue distribution of a
short diffusive junction, \cite{dorokhov1984}
\begin{align}
  \label{eq:diffusive-distribution}
  \sum_{\tau_n} = \int_0^1\dd{\tau}\rho(\tau)
  \,,
  \quad
  \rho(\tau) = \frac{1}{\tau\sqrt{1 - \tau}}
  \,.
\end{align}
Changing the integration variable to $q$ for $\beta=+$ and $q^{-1}$ for $\beta=-$ in
Eq.~\eqref{eq:stationary-statistics}:
\begin{align}
  \label{eq:W-changed}
  \ln W_R(u)
  &=
  2t_0
  \int_{0}^\infty\frac{\dd{E}}{2\pi}
  \int_0^\infty\frac{\dd{q}}{q}
  \tau(q)\sqrt{1 - \tau(q)} \rho(\tau(q))
  \\
  \notag
  &\qquad\times
  \ln
  [1
  +
  \frac{4\lambda q}{(1 + \lambda q)^2}
  F(E,u)
  ]
  \,,
\end{align}
where $F(E,u) = (e^{iuE} - 1)f_L (1 - f_R) + (e^{-iuE} - 1) f_R (1 -
f_L)$.  The diffusive distribution~\eqref{eq:diffusive-distribution}
cancels the $\tau(q)$ dependent prefactors. The integral can then 
be evaluated:
\begin{align}
  \label{eq:W-diffusive}
  \ln W_R(u)
  &=
  2t_0
  \int_{\max(\Delta_L,\Delta_R)}^\infty\frac{\dd{E}}{2\pi}
  4\arcsinh^2\sqrt{F(E,u)}
  \,.
\end{align}
Note that the result does not depend on $\lambda$. The heat transport
statistics of a diffusive junction is \emph{independent} of the phase
difference between the superconductors. Moreover, the only difference
to the normal-state result is the presence of a gap
$\max(|\Delta_L|,|\Delta_R|)$ in the energy integration.  

The absence of phase oscillations arises as coherent contributions
from different channels cancel each other. The sign of the phase
oscillation is different for large and small $\tau_n$, and diffusive
junctions contain the exact balance of low and high-transparency
channels necessary for the sum to cancel. The diffusive limit
distribution is unique in the sense that $\log{}q_n$ are uniformly
distributed \cite{beenakker97}, which is crucial for the above result.

\begin{figure}
  \includegraphics{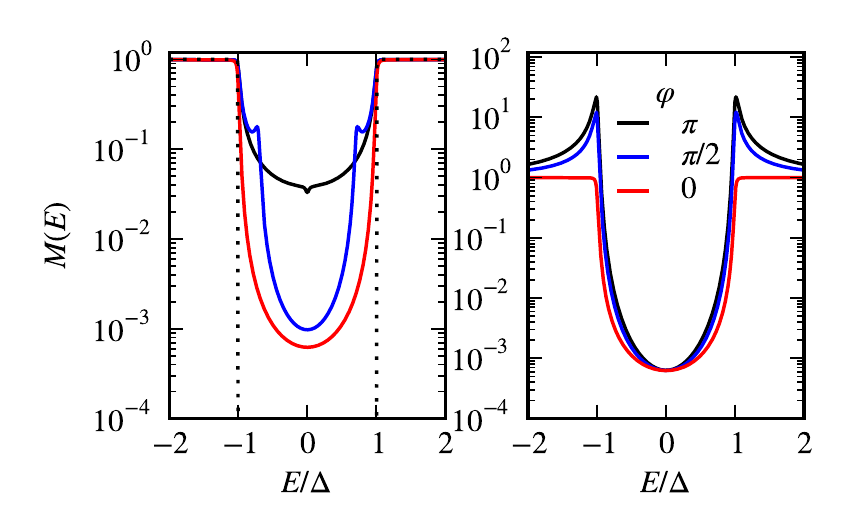}
  \caption{\label{fig:Tsum-Gamma-diffusive} Heat transparency
    of a short diffusive junction (left
    panel) and tunnel junction channel $\tau=0.01$ (right panel), for
    inelastic Dynes parameter $\Gamma=0.05\Delta$. The diffusive-limit
    phase-independent result for $\Gamma\to0^+$ is shown as a dotted
    line in the left panel.  
  }
\end{figure}

Moreover, it should be noted that the phase dependence reappears in
the diffusive limit if inelastic scattering is present
($\Gamma>0$). The technical reason for this is that it is only in the
limit $\Gamma\to0$ that superconductivity appears solely as a scale
factor $\lambda$ in the square transfer matrix eigenvalues $q$.

We can illustrate the reappearance of phase oscillations via the
normalized value $M(E)=\sum_{n,\pm}T_{n,\pm}(E) /[2\sum_n \tau_n]$ that
can be understood as a transparency for the heat
current~\eqref{eq:heat-current}. The result is shown in
Fig.~\ref{fig:Tsum-Gamma-diffusive}. At sub-gap energies, the
emergence of the well-known diffusive junction minigap of size
$E_g=|\Delta||\cos\frac{\varphi}{2}|$ is evident.  Moreover, as
pointed out above for single transparent channels, in stark contrast
with low-transparency tunnel junctions, the relative sub-gap change in
heat conductivity can be several orders of magnitude in transparent
junctions (left panel).

\subsection{Dirty interfaces}

A second universal, potentially experimentally interesting eigenvalue
distribution is the ``dirty interface'' distribution \cite{schep97,melsen1994}
\begin{align}
  \label{eq:dirty-interface-rho}
  \rho(\tau)
  =
  \frac{g}{\pi}\frac{1}{\tau^{3/2}\sqrt{1-\tau}}
  \,,
  \quad
  g = \sum_n \tau_n
  \,.
\end{align}
Experimentally, this has been found to match results existing in
high-transparency tunnel junctions. \cite{naveh2000-udt}

\begin{figure}
  \includegraphics{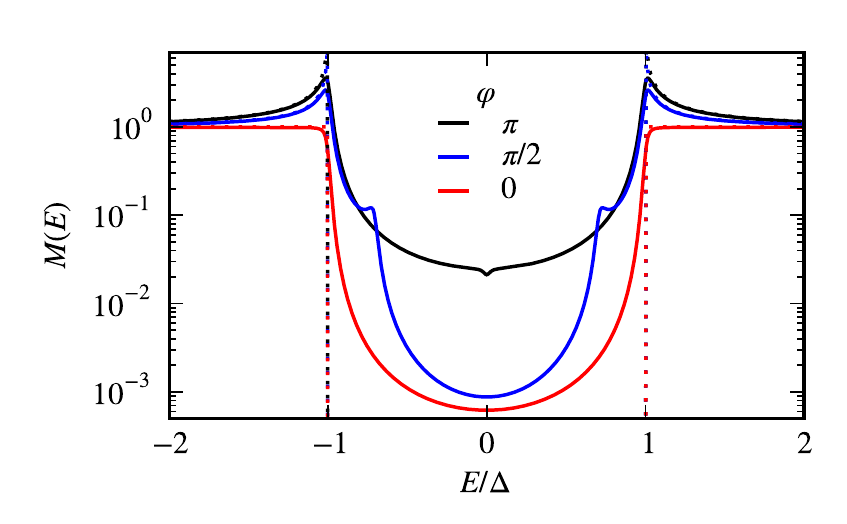}
  \caption{\label{fig:dirty}
    Heat transparency $M(E)=\sum_{n,\pm}T_{n,\pm}(E)/[2\sum_n\tau_n]$
    for the dirty interface transmission distribution.
    Shown for $\Gamma=0.05\Delta$ (solid) and $\Gamma\to0^+$ (dashed).
  }
\end{figure}

The generating function can be evaluated also in this case, starting from
Eq.~\eqref{eq:W-changed}:
\begin{align}
  \label{eq:Wdirty}
  \ln W_R(u)
  &=
  2t_0
  \int_{\max(\Delta_L,\Delta_R)}^\infty\frac{\dd{E}}{2\pi}
  4\sqrt{2}
  g
  \sinh^2\frac{z(E,u)}{4}
  \\\notag
  &\qquad\quad
  \times\sqrt{
    1
    +
    \frac{
      E^2 - \Delta_L\Delta_R\cos\varphi
    }{
      \sqrt{E^2 - \Delta_L^2}
      \sqrt{E^2 - \Delta_R^2}
    }
  }
  \,,
\end{align}
where $z(E,u)=\arccosh[1 + 2F(E,u)]$.  The associated heat transparency
entering the heat current is illustrated in Fig.~\ref{fig:dirty}.  We
can note that the above-gap transport resembles that in tunnel
junctions, and the sub-gap part that of diffusive junctions.  However,
in contrast to the tunnel junction result,
\cite{guttman1998-ieh,zhao2003-phase} the energy integral in
Eq.~\eqref{eq:Wdirty} is convergent, and does not require cutoffs in
the above-gap resonance.

\section{Discussion}

Electronic transport of heat in Josephson junctions is
facilitated by transfer of quasiparticles from one side to the other.
Andreev reflections transfer no heat, and so only quasiparticles
contribute. Consequently, the heat transport statistics follows
directly from the counting statistics of these excitations.  Exactly
as in normal-state junctions, \cite{levitov1993-cdi,levitov96} this is
determined by the transmission eigenvalues of an appropriate
scattering problem.  In the superconducting state, this statistics is
described by Bogoliubov--de Gennes transmission eigenvalues.  We find
their analytical expressions, Eqs.~\eqref{eq:tprob}
and~\eqref{eq:tfull}.

We considered both above-gap and sub-gap transport of heat, the latter
by using a toy model for the broadening of the density of states in
the superconductors.  The general picture that emerges is that sub-gap
heat transport is significantly more sensitive to the phase difference
in transparent junctions than in tunnel junctions.  Interestingly, in
diffusive junctions it is in fact only the sub-gap transport that has
any phase dependence at all.

The fluctuation statistics of heat current can be measured using
similar approaches as previously used for studying the heat currents.
\cite{giazotto2012-jhi,martinez2014-qdt} For example, one can make the heat capacity of
one of the terminals small, and then observe the fluctuation of its
total energy via temperature measurements using established
experimental techniques \cite{giazotto2006-omi}.  Setups utilizing
several metal islands \cite{heikkila2009-stf,utsumi2014-fth} could
also be a viable approach for probing the statistics experimentally.

In summary, we consider heat current driven by a temperature
difference across Josephson junctions of varying transparency.
We obtain the generating function of fluctuation statistics in closed
form. In addition to describing the fluctuations of heat current, the
results provide a way to understand the average heat current in terms
of elementary transmission events.

\acknowledgements

P.V. acknowledges the Academy of Finland for financial support.  The
work of F.G. has been partially funded by the European Research
Counsil under the European Union's Seventh Framework Programme
(FP7/2007-2013)/ERC grant agreement No. 615187-COMANCHE, and by the
Marie Curie Initial Training Action (ITN) Q-NET 264034.


\end{document}